\documentclass[12pt,a4paper]{iopart}

\usepackage{iopams}
\begin{document}

\title[Tensor-based derivation of standard vector identities]{Tensor-based derivation of standard vector identities}

\author{Miguel \'{A}ngel Rodr\'{\i}guez-Valverde and Mar\'{\i}a Tirado-Miranda}
\address{Grupo de F\'{\i}sica de Fluidos y Biocoloides, Departamento de F\'{\i}sica
Aplicada, Facultad de Ciencias, Universidad de Granada. E-18071 Granada}
\ead{marodri@ugr.es}

\begin{abstract}
Vector algebra is a powerful and needful tool for Physics but unfortunately,
due to lack of mathematical skills, it becomes misleading for first
undergraduate courses of science and engineering studies. Standard vector
identities are usually proved using Cartesian components or geometrical
arguments, accordingly. Instead, this work presents a new teaching strategy in
order to derive symbolically vector identities without analytical expansions
in components, either explicitly or using indicial notation. This strategy is
mainly based on the correspondence between three-dimensional vectors and
skew-symmetric second-rank tensors. Hence, the derivations are performed from
skew tensors and dyadic products, rather than cross products. Some
examples\ of skew-symmetric tensors in Physics are illustrated.

\end{abstract}

%\submitto{\EJP}
\maketitle

\section{Introduction}
Vector analysis \cite{murray} plays a key role in many branches of Physics:
Mechanics, Fluid dynamics, Electromagnetism theory.., because it is a powerful
mathematical tool that can express physical laws in invariant forms. Hence,
learning of vector skills must be a priority goal for science and engineering
students of undergraduate courses \cite{chyba}. However, understanding of
vectors often becomes intricate \cite{nguyen} due to the underlying
mathematics, which can even hide the meaning of the involved physical
quantities \cite{flores}. Common pitfalls are originated by the lack of
mathematical resources for deriving vector identities.

In undergraduate physics courses, the standard identities of introductory
vector algebra are mostly proved either from geometrical arguments \cite{vold}
or analytically using rectangular Cartesian components, and at best from the
indicial notation \cite{murray,chang}. Unlike the analytical proofs,
geometrical derivations are performed regardless of the coordinate system.
Instead, the demonstrations based on the indicial notation are more elegant and
compact although they require to handle complex symbolic expressions, without
any physical insight into the problem at hand.

We present an alternate approach of vector identity derivation based on the
use of tensors and dyadic products rather than \emph{cross} products. Tensor
algebra using matrix format \cite{Lass} become less cumbersome than indicial
notation and further, the operations involving second-order tensors are
readily understood as transformations of vectors.

Hereafter, only for illustrative purposes, just first- and second-rank
Cartesian tensors are considered, i.e. the three-dimensional space is
Euclidean. Hence, the contravariant and covariant components are identical to
one another because the metric tensor and conjugate metric tensor are equal to
the identity matrix. Nevertheless, the derivations compiled in this text are
equally valid for other metrices with minor modifications.

\section{Dyadics}

Aside from the well-known \emph{dot} product (particular case of the inner
product), a \emph{dyadic} is formed by the outer or direct product of two
vectors. The dyadic between the vectors ${\vec{a}}$ and ${\vec{b}}$ produces
the following second-order tensor \cite{Lass} of nine components:%
\begin{equation}
({\vec{a}\vec{b}})_{ij}\triangleq a_{i}b_{j}\label{eq:dyadic}%
\end{equation}
with $i$, $j$ $=1,2,3$ and where $a_{i}$ and $b_{j}$ are the respective
Cartesian components of both operating vectors. Unlike the inner product or
contraction, symbolized by a point, and the double inner product, symbolized
by colon, no specific symbol is employed for the dyadic product.

Since an arbitrary vector can be expressed as a linear combination of the unit
vector basis $\left\{  \hat{e}_{i}\right\}  _{i=1,2,3}$, an arbitrary dyadic
can be written into components from the concerning unit dyads $\left\{
\hat{e}_{i}\hat{e}_{j}\right\}  _{i,j=1,2,3}$ as follows:%
\begin{equation}
\vec{a}\vec{b}=\left(  {\vec{a}\vec{b}}\right)  _{ij}\hat{e}_{i}\hat{e}%
_{j}\label{eq:dyadic2}%
\end{equation}
where the summation convention is in effect for the repeated
indices \cite{murray}. If the unit vectors $\hat{e}_{i}$ are mutually
orthogonal, a special dyadic called the identical dyadic arises:%
\begin{equation}
\mathbf{1}=\hat{e}_{i}\hat{e}_{i}\label{eq:identity}%
\end{equation}
where the summation convention is again invoked. This quantity is the
second-order \emph{identity} tensor of three-dimensional space.

The inner product can be applied between vectors and second-order tensors as
well, like a matrix product keeping their own properties. Thus, dyadics hold
the following properties (derivation not shown):

\begin{itemize}
\item $\vec{a}\vec{b}=\left(  {\vec{b}\vec{a}}\right)  ^{t}$

\item $\left(  {\vec{c}\vec{a}}\right)  \cdot\vec{b}=\left(  {\vec{a}\cdot
\vec{b}}\right)  \vec{c}$

\item $\vec{c}\cdot\left(  {\vec{a}\vec{b}}\right)  =\left(  {\vec{c}\cdot
\vec{a}}\right)  \vec{b}$

\item $\left(  {\vec{a}\vec{b}}\right)  \cdot\left(  {\vec{c}\vec{d}}\right)
=\left(  {\vec{b}\cdot\vec{c}}\right)  \left(  {\vec{a}\vec{d}}\right)  $
\end{itemize}

where the superscript $t$ stands for the matrix transpose. Note that even
though the vector transpose is represented by a $1\times3$ matrix instead of
the conventional $3\times1$ matrix, the vector after transposition remains
identical, i.e. $\vec{a}\equiv\left(  {\vec{a}}\right)  ^{t}$. By default,
vectors at left-hand side in an inner product are transpose.

Although it is not used in this paper, the trace of ${\vec{a}\vec{b}}$, i.e.
the sum of their diagonal components, is indeed the concerning dot product:%
\[
\textnormal{trace}\left(  {\vec{a}\vec{b}}\right)  =\vec{a}\cdot\vec{b}%
\]
In fact, the trace of ${\vec{a}\vec{b}}$ can be expressed in terms of double
inner product as $\frac{1}{3}\vec{a}\vec{b}:\mathbf{1}$.

\section{Skew-symmetric tensor associated to a vector}
In vector algebra \cite{leubner}, the skew-symmetric tensor
$\mathbf{\Omega}{_{\vec{a}}}$ of rank two associated to a vector
${\vec{a}}$ is defined by:%
\begin{equation}
\left(\mathbf{\Omega}{_{\vec{a}}}\right)  _{ij}\triangleq
-\varepsilon_{ijk}a_{k}\label{eq:skewsymmetric}%
\end{equation}
where $\varepsilon_{ijk}$ stands for the Levi-Civita symbol \cite{Lass}, also
referred to as $\varepsilon$-permutation symbol, and where all indices have
the range 1, 2, 3. The index $k$ is the dummy summation index according to the
summation convention. The epsilon symbol $\varepsilon_{ijk}$ holds the
following rules:

\begin{itemize}
\item ${\varepsilon_{123}=\varepsilon_{231}=\varepsilon_{321}=1}$

\item ${\varepsilon_{123}=-\varepsilon_{213}=-\varepsilon_{132}}$

\item ${\varepsilon_{ijk}=0,}$ otherwise
\end{itemize}

There is an additional relation known as epsilon-delta identity:%
\begin{equation}
{\varepsilon_{mni}\varepsilon}_{ijk}{=\delta_{mj}\delta_{nk}-\delta_{mk}%
\delta_{nj}}\label{eq:epsilondelta}%
\end{equation}
where ${\delta_{ij}}$ is the Kronecker delta ($ij$-component of the
second-order identity tensor) and the summation is performed over the $i$
index. Indeed, the epsilon symbol and the Kronecker delta are both numerical
tensors which have fixed components in every coordinate system. As the
identity tensor,$\mathbf{1}$, can be generated from the summation
of the unit dyads (\ref{eq:identity}) built by any orthonormal vector basis
$\left\{  \hat{e}_{i}\right\}  _{i=1,2,3}$, the epsilon symbol can be
accordingly found from the following triple scalar product:%
\[
\varepsilon_{ijk}=\left(  \hat{e}_{i}\times\hat{e}_{j}\right)  \cdot\hat
{e}_{k}%
\]
where the cross product is symbolized by $\times$. From the anti-cyclic rule
of $\varepsilon_{ijk}$ and the definition (\ref{eq:skewsymmetric}), it is
straightforwardly shown that the tensor $\mathbf{\Omega}_{\vec{a}%
}$ is anti-symmetric:%
\begin{equation}
\mathbf{\Omega}_{\vec{a}}^{t}=-\mathbf{\Omega}%
_{\vec{a}}\label{eq:skewsymmetry}%
\end{equation}
and this can be readily illustrated from the matrix form of
$\mathbf{\Omega}_{\vec{a}}$:%
\[
\mathbf{\Omega}_{\vec{a}}=\left(
\begin{array}
[c]{ccc}%
0 & -a_{3} & a_{2}\\
a_{3} & 0 & -a_{1}\\
-a_{2} & a_{1} & 0
\end{array}
\right)
\]
Also, $\vec{a}$ is called the (Hodge) dual vector of the skew-symmetric tensor
$\mathbf{\Omega}_{\vec{a}}$. Hence, for instance, the magnetic
field tensor in Electrodynamics \cite{roche} is indeed the skew-symmetric
tensor associated to the magnetic field vector.

The Levi-Civita symbol also appears in the definition of the cross product of
${\vec{a}}$ and ${\vec{b}}$ \cite{murray}:%
\begin{equation}
\left(  {\vec{a}\times\vec{b}}\right)  _{i}\triangleq\varepsilon_{ijk}%
a_{j}b_{k}%
\end{equation}
then, from the definition (\ref{eq:skewsymmetric}) and the anti-cyclic rule of
the epsilon symbol, it is possible rewritten the cross product in terms of the
concerning skew-symmetric tensor (\ref{eq:skewsymmetric}) as:
\begin{equation}
\left(  {\vec{a}\times\vec{b}}\right)  _{i}=\left(  \mathbf
{\Omega}{_{\vec{a}}}\right)  _{ik}b_{k}%
\end{equation}
or in vector notation as:%
\begin{equation}
\vec{a}\times\vec{b}=\mathbf{\Omega}_{\vec{a}}\cdot\vec
{b}\label{eq:crossproduct}%
\end{equation}

A cross product typically returns a (true) vector or polar vector. More
exactly, the cross product (\ref{eq:crossproduct}) is a vector if either
${\vec{a}}$ or ${\vec{b}}$ (but not both) are pseudovectors. Otherwise,
$\vec{a}\times\vec{b}$ is a pseudovector \cite{hauser}. Then, it is worthy to
mention that the tensor $\mathbf{\Omega}_{\vec{a}}$ will be a
relative tensor or pseudotensor if the vector ${\vec{a}}$ is axial and
otherwise, it will be an absolute tensor if the vector ${\vec{a}}$ is polar.

In addition to the skew-symmetry (\ref{eq:skewsymmetry}), the tensor
$\mathbf{\Omega}_{\vec{a}}$ holds the following properties
(derivation not shown):

\begin{itemize}
\item $\mathbf{\Omega}{_{\alpha\vec{a}}=\alpha}\mathbf
{\Omega}_{\vec{a}}$

\item $\mathbf{\Omega}{_{\vec{a}+\vec{b}}=}\mathbf
{\Omega}_{\vec{a}}+\mathbf{\Omega}_{\vec{b}}$

\item $\mathbf{\Omega}{_{\vec{a}}\cdot\vec{a}=\vec{0}}$

\item $\mathbf{\Omega}{_{\vec{b}}\cdot\vec{a}=\vec{b}%
\cdot\mathbf{\Omega}_{\vec{a}}}$

\item $\mathbf{\Omega}{_{\vec{a}}\cdot\mathbf{\Omega
}_{\vec{b}}=\vec{b}\vec{a}-\left(  {\vec{a}\cdot\vec{b}}\right)
}\mathbf{{1}}$

\item $\mathbf{\Omega}{_{\vec{a}\times\vec{b}}=\vec{b}\vec{a}%
-\vec{a}\vec{b}=\mathbf{\Omega}_{\vec{a}}\cdot\mathbf
{\Omega}_{\vec{b}}-\mathbf{\Omega}_{\vec{b}}\cdot
\mathbf{\Omega}_{\vec{a}}}$
\end{itemize}

where $\alpha$ is a scalar. These properties can be straightforwardly proved
using index notation and the above-mentioned rules of the Levi-Civita symbol.
Thus, the epsilon-delta identity (\ref{eq:epsilondelta}) draws to the last two
properties, which are very helpful for the derivations compiled in section
\ref{sec:standard vector identities}. In particular, these other properties
are also very useful:

\begin{itemize}
\item ${\mathbf{\Omega}_{-\vec{a}}=\mathbf{\Omega
}_{\vec{a}}^{t}}$

\item $\mathbf{\Omega}{_{\vec{a}}\cdot\mathbf{\Omega
}_{\vec{b}}=}\left(  \mathbf{\Omega}{_{\vec{b}}\cdot
\mathbf{\Omega}_{\vec{a}}}\right)  ^{t}$

\item ${\mathbf{\Omega}_{\vec{a}}\cdot\vec{b}=-\vec{b}%
\cdot\mathbf{\Omega}_{\vec{a}}}$

\item ${\mathbf{\Omega}_{\hat{e}}^{2}=\hat{e}\hat{e}%
-}\mathbf{{1}}$

\item ${\mathbf{\Omega}_{\hat{e}}^{3}=-\mathbf{\Omega
}_{\hat{e}}}$
\end{itemize}

where ${\hat{e}}$ is a vector of unit length. Due to Eq. (\ref{eq:identity}),
${\mathbf{\Omega}_{\hat{e}}^{2}}$ is equal to the second-order
identity tensor of the two-dimensional space (plane) with normal unit
${\hat{e}}$.

\section{\label{sec:standard vector identities}Standard vector identities}
Next, the most useful vector identities are demostrated from the concerning
dyadics (\ref{eq:dyadic}) and skew-symmetric tensors (\ref{eq:skewsymmetric}).
The above-listed properties, the associative rule of matrix product and the
matrix transposition rules are used accordingly.

\begin{itemize}
\item Cyclic permutation of the scalar triple product:%
\begin{eqnarray}
\left(  {\vec{a}\times\vec{b}}\right)  \cdot\vec{c}  &  =\left(  {\vec{a}%
\cdot\mathbf{\Omega}_{\vec{b}}}\right)  \cdot\vec{c}=\vec{a}%
\cdot\left(  \mathbf{\Omega}{_{\vec{b}}\cdot\vec{c}}\right)
=\vec{a}\cdot\left(  {\vec{b}\times\vec{c}}\right) \nonumber\\
&  =\left(  {\vec{b}\cdot\mathbf{\Omega}_{-\vec{a}}}\right)
\cdot\vec{c}=\vec{b}\cdot\left(  \mathbf{\Omega}{_{-\vec{a}}%
\cdot\vec{c}}\right)  =\vec{b}\cdot\left(  {\vec{c}\times\vec{a}}\right)
\label{eq:scalartripleproduct}%
\end{eqnarray}
From these identities, the orthogonality between ${\vec{a}\times\vec{b}}$ and
each vector can be readily illustrated:%
\begin{equation}
\left(  {\vec{a}\times\vec{b}}\right)  \cdot{\vec{a}=}\vec{0}
\label{eq:scalartripleproduct2}%
\end{equation}

\item Vector triple product expansion (or Lagrange's formula):%
\begin{eqnarray}
\vec{a}\times\left(  {\vec{b}\times\vec{c}}\right)   &  =\vec{a}%
\cdot\mathbf{\Omega}_{\vec{b}\times\vec{c}}=\vec{a}\cdot\left(
{\vec{c}\vec{b}-\vec{b}\vec{c}}\right)  \nonumber\\
&  =\left(  {\vec{a}\cdot\vec{c}}\right)  \vec{b}-\left(  {\vec{a}\cdot\vec
{b}}\right)  \vec{c}\label{eq:Lagrange}%
\end{eqnarray}
also known as the acb minus abc rule. Using this identity, any vector can be
expressed as linear combination of two mutually perpendicular vectors
according to an arbitrary direction, ${\hat{e}}$:%
\begin{equation}
\vec{a}=\left(  {\vec{a}\cdot\hat{e}}\right)  {\hat{e}+\hat{e}}\times\left(
\vec{a}{\times\hat{e}}\right)  \label{eq:perpendiculardescomposicion}%
\end{equation}
Furthermore, Eq. (\ref{eq:Lagrange}) might be used directly in many
below-mentioned identities, thereby it is one of the most important vector identities.

\item Jacobi's identity:%
\begin{eqnarray}
\vec{a}\times\left(  {\vec{b}\times\vec{c}}\right)  +\vec{b}\times\left(
{\vec{c}\times\vec{a}}\right)  +\vec{c}\times\left(  {\vec{a}\times\vec{b}%
}\right)   &  =\vec{a}\cdot\mathbf{\Omega}_{\vec{b}\times\vec{c}%
}+\vec{b}\cdot\mathbf{\Omega}_{\vec{c}\times\vec{a}}+\nonumber\\
+\vec{c}\cdot\mathbf{\Omega}_{\vec{a}\times\vec{b}} &  =\vec
{a}\cdot\left(  {\vec{c}\vec{b}-\vec{b}\vec{c}}\right)  +\nonumber\\
+\vec{b}\cdot\left(  {\vec{a}\vec{c}-\vec{c}\vec{a}}\right)  +\vec{c}%
\cdot\left(  {\vec{b}\vec{a}-\vec{a}\vec{b}}\right)   &  =\vec{0}%
\label{eq:Jacobi}%
\end{eqnarray}
i.e. the sum of all the cyclic permutations of the vector double product comes
to zero.

\item Dot product of two cross products:%
\begin{eqnarray}
\left(  {\vec{a}\times\vec{b}}\right)  \cdot\left(  {\vec{c}\times\vec{d}%
}\right)   &  =\left(  {\mathbf{\Omega}_{\vec{a}}\cdot\vec{b}%
}\right)  \cdot\left(  {\mathbf{\Omega}_{\vec{c}}\cdot\vec{d}%
}\right)  =\vec{b}\cdot\left(  {\mathbf{\Omega}_{-\vec{a}}%
\cdot\mathbf{\Omega}_{\vec{c}}}\right)  \cdot\vec{d}\nonumber\\
&  =\vec{b}\cdot\left(  {\left(  {\vec{a}\cdot\vec{c}}\right)
\mathbf{1}-\vec{c}\vec{a}}\right)  \cdot\vec{d}\nonumber\\
&  =\vec{b}\cdot\left(  {\left(  {\vec{a}\cdot\vec{c}}\right)  \vec{d}-\left(
{\vec{a}\cdot\vec{d}}\right)  \vec{c}}\right) \nonumber\\
&  =\left(  {\vec{a}\cdot\vec{c}}\right)  \left(  {\vec{b}\cdot\vec{d}%
}\right)  -\left(  {\vec{a}\cdot\vec{d}}\right)  \left(  {\vec{b}\cdot\vec{c}%
}\right)  \label{eq:doublescalarcross}%
\end{eqnarray}

This identity and the next one afford a simple means of deducing the formulae
of Spherical Trigonometry. From Eq. (\ref{eq:doublescalarcross}), it is also
derived the well-known identity:
\begin{equation}
\left(  {\vec{a}\times\vec{b}}\right)  ^{2}=\left(  {\vec{a}\cdot\vec{a}%
}\right)  \left(  {\vec{b}\cdot\vec{b}}\right)  -\left(  {\vec{a}\cdot\vec{b}%
}\right)  ^{2} \label{eq:cuadrados}%
\end{equation}
which geometrical interpretation is the Pythagorean theorem.

\item Cross product of two cross products:%
\begin{eqnarray}
\left(  {\vec{a}\times\vec{b}}\right)  \times\left(  {\vec{c}\times\vec{d}%
}\right)   &  ={\mathbf{\Omega}}_{\vec{a}\times\vec{b}}%
\cdot\left(  {\mathbf{\Omega}_{\vec{c}}\cdot\vec{d}}\right)
=\left(  {\vec{b}\vec{a}-\vec{a}\vec{b}}\right)  \cdot\left(
{\mathbf{\Omega}_{\vec{c}}\cdot\vec{d}}\right) \nonumber\\
&  =\left(  {\vec{b}\vec{a}}\right)  \cdot\left(  {\mathbf{\Omega
}_{\vec{c}}\cdot\vec{d}}\right)  -\left(  {\vec{a}\vec{b}}\right)
\cdot\left(  {\mathbf{\Omega}_{\vec{c}}\cdot\vec{d}}\right)
\nonumber\\
&  =\left(  {\vec{a}\cdot\left(  {\mathbf{\Omega}_{\vec{c}}%
\cdot\vec{d}}\right)  }\right)  \vec{b}-\left(  {\vec{b}\cdot\left(
{\mathbf{\Omega}_{\vec{c}}\cdot\vec{d}}\right)  }\right)  \vec
{a}\nonumber\\
&  =\left(  {\vec{a}\cdot\left(  {\vec{c}\times\vec{d}}\right)  }\right)
\vec{b}-\left(  {\vec{b}\cdot\left(  {\vec{c}\times\vec{d}}\right)  }\right)
\vec{a}%
\end{eqnarray}%
\begin{eqnarray}
\left(  {\vec{a}\times\vec{b}}\right)  \times\left(  {\vec{c}\times\vec{d}%
}\right)   &  =\left(  {\vec{a}\cdot\mathbf{\Omega}_{\vec{b}}%
}\right)  \cdot{\mathbf{\Omega}}_{\vec{c}\times\vec{d}}=\left(
{\vec{a}\cdot\mathbf{\Omega}_{\vec{b}}}\right)  \cdot\left(
{\vec{d}\vec{c}-\vec{c}\vec{d}}\right) \nonumber\\
&  =\left(  {\vec{a}\cdot\mathbf{\Omega}_{\vec{b}}}\right)
\cdot\left(  {\vec{d}\vec{c}}\right)  -\left(  {\vec{a}\cdot
\mathbf{\Omega}_{\vec{b}}}\right)  \cdot\left(  {\vec{c}\vec{d}%
}\right) \nonumber\\
&  =\left(  {\vec{a}\cdot\left(  {\mathbf{\Omega}_{\vec{b}}%
\cdot\vec{d}}\right)  }\right)  \vec{c}-\left(  {\vec{a}\cdot\left(
{\mathbf{\Omega}_{\vec{b}}\cdot\vec{c}}\right)  }\right)  \vec
{d}\nonumber\\
&  =\left(  {\vec{a}\cdot\left(  {\vec{b}\times\vec{d}}\right)  }\right)
\vec{c}-\left(  {\vec{a}\cdot\left(  {\vec{b}\times\vec{c}}\right)  }\right)
\vec{d}%
\end{eqnarray}

\item Other identities:%
\begin{eqnarray}
\left(  {\left(  {\vec{a}\times\vec{b}}\right)  \times\vec{b}}\right)
\cdot\vec{a}  &  =\left(  {\mathbf{\Omega}_{\vec{a}\times\vec{b}%
}\cdot\vec{b}}\right)  \cdot\vec{a}=\left(  {\left(  {\vec{b}\vec{a}-\vec
{a}\vec{b}}\right)  \cdot\vec{b}}\right)  \cdot\vec{a}\nonumber\\
&  =\left(  {\left(  {\vec{a}\cdot\vec{b}}\right)  \vec{b}-\left(  {\vec
{b}\cdot\vec{b}}\right)  \vec{a}}\right)  \cdot\vec{a}\nonumber\\
&  =-\left(  {\vec{a}\times\vec{b}}\right)  ^{2}%
\end{eqnarray}
The identity (\ref{eq:cuadrados}) was invoked for this result. However,
$\left(  {\left(  {\vec{a}\times\vec{b}}\right)  \times\vec{b}}\right)
\cdot{\vec{b}}=0$ is readily found from Eq. (\ref{eq:scalartripleproduct2}).
Likewise, in the next two expressions, the identity
(\ref{eq:scalartripleproduct2}) was again applied accordingly:%
\begin{eqnarray}
\left(  {\left(  {\vec{a}\times\vec{b}}\right)  \times\vec{b}}\right)
\times\vec{a}  &  ={\mathbf{\Omega}}_{\left(  {\vec{a}\times
\vec{b}}\right)  \times\vec{b}}\cdot\vec{a}=\left(  {\vec{b}\left(  {\vec
{a}\times\vec{b}}\right)  -\left(  {\vec{a}\times\vec{b}}\right)  \vec{b}%
}\right)  \cdot\vec{a}\nonumber\\
&  =-\left(  {\vec{a}\cdot\vec{b}}\right)  \left(  {\vec{a}\times\vec{b}%
}\right) \\
\left(  {\left(  {\vec{a}\times\vec{b}}\right)  \times\vec{b}}\right)
\times\vec{b}  &  ={\mathbf{\Omega}}_{\left(  {\vec{a}\times
\vec{b}}\right)  \times\vec{b}}\cdot\vec{b}=\left(  {\vec{b}\left(  {\vec
{a}\times\vec{b}}\right)  -\left(  {\vec{a}\times\vec{b}}\right)  \vec{b}%
}\right)  \cdot\vec{b}\nonumber\\
&  =-\left(  {\vec{b}\cdot\vec{b}}\right)  \left(  {\vec{a}\times\vec{b}%
}\right)
\end{eqnarray}

\end{itemize}
\section{Skew-symmetric tensors in Physics}

The substitution of physical pseudovectors (such as angular velocity or
magnetic field) with skew-symmetric tensors (\ref{eq:skewsymmetric}) provides
an alternate to cross product. This notation is much easier to work and allows
to understand the vector operations in terms of rotations \cite{koehler}. In
fact, an arbitrary vector ${\vec{a}}$, which rotates an angle $\theta$ about
an axis along the unit vector ${\hat{e}}$ \cite{mathews}, is expressed by:%
\begin{equation}
{\vec{a}}^{\ast}=\left(  \mathbf{{1}}+\left(  \left(  1-\cos
\theta\right)  {\mathbf{\Omega}_{\hat{e}}+\sin\theta
}\mathbf{{1}}\right)  \cdot{\mathbf{\Omega}}_{{\hat
{e}}}\right)  \cdot{\vec{a}}%
\end{equation}

\subsection{Rotating systems}

A system rotates with constant angular velocity, $\vec{\omega}$, relative to a
rest frame. The time variation of a unit vector $\hat{e}$ fixed to the
rotating system \cite{marion,weinstock} is given by:%
\begin{equation}
\frac{{d\hat{e}}}{{dt}}=\vec{\omega}\times\hat{e}={\mathbf{\Omega
}}_{\vec{\omega}}\cdot\hat{e}%
\end{equation}
where the tensor ${\mathbf{\Omega}}_{\vec{\omega}}$ acts as a
rotation operator. Also, if a point particle linked to the rotating system is
moving at linear velocity, $\vec{v}$, regarding to the rest frame, then it
undergoes the following Coriolis' acceleration \cite{marion,weinstock}:%
\begin{equation}
\vec{a}_{Coriolis}=2\vec{\omega}\times\vec{v}={\mathbf{\Omega}%
}_{2\vec{\omega}}\cdot\vec{v}%
\end{equation}
where now the rotated vector is $\vec{v}$.

\subsection{Rotation dynamics of rigid body motion}

A system of $N-$point particles of mass ${m_{i}}$ and position vector
${\vec{r}_{i}}$ relative to a rest reference frame, describes a pure rotation
with angular velocity ${\vec{\omega}}$ and acceleration ${\vec{\alpha}}%
$ \cite{marion}.

\textbullet Kinematics

The linear velocity and acceleration of the $i$-particle are rewritten as:%
\begin{equation}
\vec{v}_{i}=\vec{\omega}\times\vec{r}_{i}={\mathbf{\Omega}}%
_{\vec{\omega}}\cdot\vec{r}_{i}%
\end{equation}%
\begin{equation}
\vec{a}_{i}=\vec{\alpha}\times\vec{r}_{i}+\vec{\omega}\times\left(
{\vec{\omega}\times\vec{r}_{i}}\right)  =\left(  {\mathbf{\Omega
}_{\vec{\alpha}}+\mathbf{\Omega}_{\vec{\omega}}^{2}}\right)
\cdot\vec{r}_{i}%
\end{equation}

\textbullet Inertia tensor

The inertia tensor relative to the rest coordinate system is given by:
\begin{equation}
\mathbf{I}\triangleq\sum\limits_{i=1}^{N}{m_{i}\left(  r_{i}%
^{2}{\mathbf{1}-\vec{r}_{i}\vec{r}_{i}}\right)  }=\sum
\limits_{i=1}^{N}{m}_{i}{\mathbf{\Omega}_{\vec{r}_{i}}%
\cdot\mathbf{\Omega}_{-\vec{r}_{i}}}%
\end{equation}

\textbullet Steiner's theorem

The inertia tensor relative to a second rest frame is:%
\begin{equation}
\mathbf{I}^{\ast}=\mathbf{I}-\left(  {\sum
\limits_{i=1}^{N}{m_{i}}}\right)  \mathbf{\Omega}_{\vec{r}_{c}%
}\cdot\mathbf{\Omega}_{-\vec{r}_{c}}%
\end{equation}
where $\vec{r}_{c}$ is the position vector of the system center regarding to
the initial rest coordinate system.

\textbullet Inertia momentum with respect to an axis in the direction
${\hat{e}}$%
\begin{equation}
I_{\hat{e}}\triangleq\sum\limits_{i=1}^{N}{m_{i}\left(  {\hat{e}\times\vec
{r}_{i}}\right)  ^{2}}=\sum\limits_{i=1}^{N}{m_{i}\left(  {\hat{e}%
\cdot\mathbf{\Omega}}_{{\vec{r}_{i}}}\right)  \cdot\left(
{\mathbf{\Omega}}_{-{\vec{r}_{i}}}{\cdot\hat{e}}\right)  }=\hat
{e}\cdot\mathbf{I}\cdot\hat{e}%
\end{equation}

\textbullet Angular momentum%
\begin{equation}
\vec{L}\triangleq\sum\limits_{i=1}^{N}{m_{i}\vec{r}_{i}\times\left(
{\vec{\omega}\times\vec{r}_{i}}\right)  }=\sum\limits_{i=1}^{N}{m_{i}%
\mathbf{\Omega}}_{{\vec{r}_{i}}}{\cdot\left(  {\mathbf
{\Omega}}_{-{\vec{r}_{i}}}{\cdot\vec{\omega}}\right)  }=\sum\limits_{i=1}%
^{N}{m_{i}\left(  {\mathbf{\Omega}}_{{\vec{r}_{i}}}{\cdot
\mathbf{\Omega}}_{-{\vec{r}_{i}}}\right)  \cdot\vec{\omega}%
}=\mathbf{I}\cdot\vec{\omega}%
\end{equation}

\textbullet Rotation kinetic energy%
\begin{equation}
E_{k}\triangleq\frac{1}{2}\sum\limits_{i=1}^{N}{m_{i}\left(  {\vec{\omega
}\times\vec{r}_{i}}\right)  ^{2}}=\frac{1}{2}\sum\limits_{i=1}^{N}%
{m_{i}\left(  {\vec{\omega}\cdot\mathbf{\Omega}}_{{\vec{r}_{i}}%
}\right)  \cdot\left(  {\mathbf{\Omega}}_{-{\vec{r}_{i}}}%
{\cdot\vec{\omega}}\right)  }=\frac{1}{2}\vec{\omega}\cdot\mathbf
{I}\cdot\vec{\omega}%
\end{equation}

\subsection{Electric quadrupole}

The quadrupole moment tensor of a system of point electric charges $\left\{
q_{i}\right\}  _{i=1..N}$ can be expressed as:%
\begin{equation}
\mathbf{Q}\triangleq\sum\limits_{i=1}^{N}{q_{i}\left(  3{\vec
{r}_{i}\vec{r}_{i}-}r_{i}^{2}{\mathbf{1}}\right)  }=3\sum
\limits_{i=1}^{N}{q}_{i}{\mathbf{\Omega}_{\vec{r}_{i}}^{2}%
+2}\left(  \sum\limits_{i=1}^{N}{q_{i}}r_{i}^{2}\right)  {\mathbf
{1}}%
\end{equation}
This second-rank tensor is traceless.

\subsection{Vector field identities}

Vector field identities can be also derived using the skew-symmetric tensor
associated to the differential vector operator $\nabla$ rather than the curl 
\cite{wimmel}, as follows:%
\begin{equation}
\nabla\times\vec{a}={\mathbf{\Omega}}_{\nabla}\cdot\vec{a}%
\end{equation}
However, special care must be taken\ in tensor calculus because the order of
elements is important:

\begin{itemize}
\item $\left(  {\nabla\vec{a}}\right)  ^{t}\neq\vec{a}\nabla$

\item $\vec{a}\cdot{\mathbf{\Omega}}_{\nabla}\neq
-{\mathbf{\Omega}}_{\nabla}\cdot\vec{a}$

\item $\left(  {\mathbf{\Omega}_{\vec{a}}\cdot\mathbf
{\Omega}_{\nabla}}\right)  ^{t}\neq{\mathbf{\Omega}}_{\nabla}%
\cdot{\mathbf{\Omega}}_{\vec{a}}$

\item $\left(  {\vec a\nabla} \right)  \cdot\vec b = \left(  {\nabla\cdot\vec
b} \right)  \vec a$

\item $\left(  {\nabla\vec a} \right)  ^{t} \cdot\vec b = \left(  {\vec b
\cdot\nabla} \right)  \vec a$
\end{itemize}

Thus, the most relevant properties are:

\begin{itemize}
\item ${\mathbf{\Omega}}_{\nabla}\cdot{\mathbf{\Omega
}}_{\vec{a}}=\left(  {\nabla\vec{a}}\right)  ^{t}-\left(  {\nabla\cdot\vec{a}%
}\right)  {\mathbf{1}}$

\item ${\mathbf{\Omega}}_{\vec{a}}\cdot{\mathbf
{\Omega}}_{\nabla}=\left(  {\vec{a}\nabla}\right)  ^{t}-{\mathbf
{1}}\left(  {\vec{a}\cdot\nabla}\right)  $

\item ${\mathbf{\Omega}}_{\nabla}\cdot\left(  {\mathbf
{\Omega}_{\vec{a}}\cdot\vec{b}}\right)  =\left(  {\mathbf{\Omega
}_{\nabla}\cdot\mathbf{\Omega}_{\vec{a}}}\right)  \cdot\vec
{b}+\left(  {\mathbf{\Omega}_{\vec{a}}\cdot\mathbf
{\Omega}_{\nabla}}\right)  ^{t}\cdot\vec{b}$

\item $\nabla\cdot\left(  {\mathbf{\Omega}_{\vec{a}}\cdot\vec{b}%
}\right)  =\left(  {\mathbf{\Omega}_{\vec{b}}\cdot\nabla}\right)
\cdot\vec{a}+\left(  {\mathbf{\Omega}_{\vec{a}}\cdot\nabla
}\right)  ^{t}\cdot\vec{b}$
\end{itemize}

\ack
This work was supported by the "Ministerio Espa\~{n}ol de Educaci\'{o}n y
Ciencia" (contract "Ram\'{o}n y Cajal" RYC-2005-000983), the European Social
Fund (ESF) and the "Junta de Andalucia" (project FQM-02517).

\newpage

\end{document}